\documentclass[aps,prl,reprint,groupedaddress]{revtex4-1}
\usepackage{amssymb}
\usepackage{color}
\usepackage{graphicx}
\usepackage{epsfig}
\usepackage{epstopdf}

\def\XXint#1#2#3{{\setbox0=\hbox{$#1{#2#3}{\int}$ }
\vcenter{\hbox{$#2#3$ }}\kern-.6\wd0}}

\newcommand{\grl}{    {Geophys. Res. Lett.}}
\newcommand{\jgr}{    {J. Geophys. Res.}}
\newcommand{\ssr}{    {Space Sci. Rev.}}

\begin{document}


\title{Electrostatic steepening of whistler waves}


\author{I.Y. Vasko$^{1}$}
\email[]{ivan.vasko.ssl@berkeley.edu}
\author{O.V. Agapitov$^{1,3}$}
\author{F.S. Mozer$^1$}
\author{J.W. Bonnell$^{1}$}
\author{A.V. Artemyev$^{4,2}$}
\author{V.V. Krasnoselskikh$^5$}
\author{Y. Tong$^1$}


\affiliation{$^1$ Space Sciences Laboratory, University of California, Berkeley, USA}
\affiliation{$^2$ Space Research Institute RAS, Moscow, Russia}
\affiliation{$^3$ National Taras Shevchenko University of Kyiv, Ukraine}
\affiliation{$^4$ University of California in Los Angeles, Los Angeles, USA}
\affiliation{$^5$ LPC2E, University of Orleans, France}



\date{\today}

\begin{abstract}
We present surprising observations by the NASA Van Allen Probes spacecraft of whistler waves with substantial electric field power at harmonics of the whistler wave fundamental frequency. The wave power at harmonics is due to nonlinearly steepened whistler electrostatic field that becomes possible in the two-temperature electron plasma due to whistler wave coupling to the electron-acoustic mode. The simulation and analytical estimates show that the steepening takes a few tens of milliseconds. The hydrodynamic energy cascade to higher frequencies facilitates efficient energy transfer from cyclotron resonant electrons, driving the whistler waves, to lower energy electrons.
\end{abstract}


\maketitle

Whistler waves play fundamental role in electron acceleration in space \citep{Thorne13,Mozer14}, solar wind \citep{Gary12} and astrophysical \citep{Petrosian04} plasmas and continuously stimulate laboratory plasma experiments \citep{VanCompernolle14,VanCompernolle15}. In particular, whistler waves control the dynamics of the Van Allen radiation belts \citep{Kennel&Petschek66}, where they are generated via the cyclotron resonant instability with regularly injected $\sim 10$ keV anisotropic electrons \citep[][]{Omura13:AGU}. Whistler waves mediate the energy of the injected electrons to higher and lower energy electrons via the resonant interaction resulting in electron acceleration up to relativistic energies \citep{Thorne13,Mozer14} and electron losses to the atmosphere \citep{Thorne10:nat}.

Whistler waves are typically observed in the form of quasi-monochromatic wave packets propagating quasi-parallel to the background magnetic field \citep{Santolik03}. The whistler wave field is decomposed into the electrostatic field along the wave vector ${\bf k}$ and elliptically polarized electromagnetic field perpendicular to ${\bf k}$ \citep{Helliwell65}. Whistler waves are fundamentally different from compressible sound waves in fluids and plasmas \citep{Landau59,Sagdeev66} in that the whistler electrostatic field has been argued not to steepen \citep{Yoon14}. In accordance, the reported whistler waves in the Van Allen radiation belts typically have quasi-sinusoidal waveforms even at the highest observed amplitudes \citep{Cattell08}. Slightly non-sinusoidal waveforms have been attributed to electrons trapped within the whistler electrostatic field \citep{Kellogg10}.

In this Letter we present whistler waves with surprisingly significant electric field power at harmonics of the fundamental frequency that is due to highly non-sinusoidal waveform of the whistler electrostatic field. The wave energy cascade to higher frequencies is due to the classical hydrodynamic steepening that becomes possible in the two-temperature electron plasma.

The twin NASA Van Allen Probe spacecraft launched on August 30, 2012 into the Van Allen radiation belts provide wave and particle measurements with unprecedent time resolution. We present Van Allen Probe A measurements on May 1, 2013 near the Earth magnetic dipole equator at the geocentric radial distance of 5.5 Earth radii. During the considered time interval the background magnetic field and electron density are about 87 nT and 3.2 cm$^{-3}$. The electron cyclotron and plasma frequencies are $f_{c}\sim 2.5$ kHz and $f_{p}\sim 16$ kHz.

\begin{figure*}
\includegraphics[width=40pc]{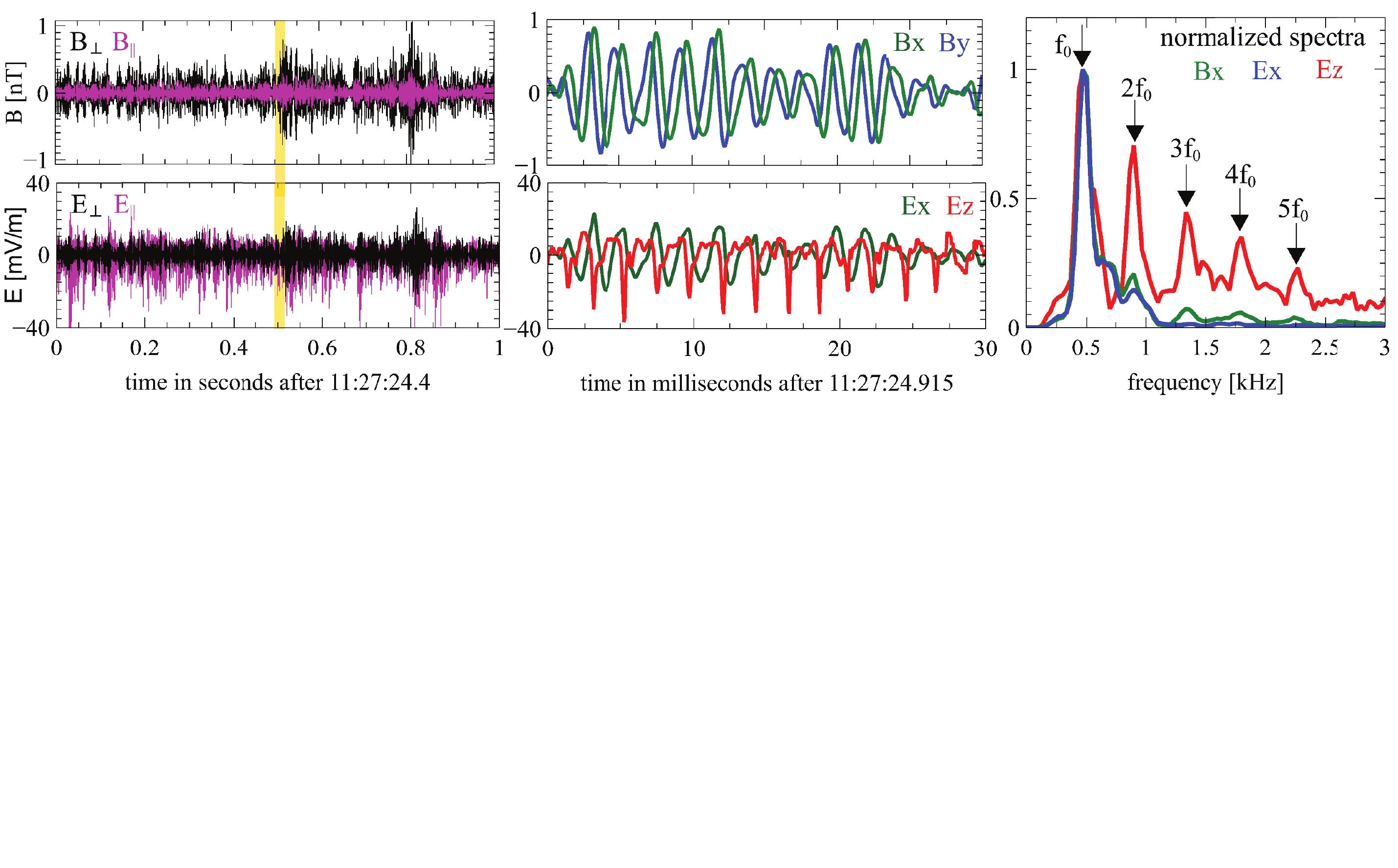}
\caption{The left panels present electric and magnetic field waveforms measured with cadence 16384 samples/s by the Electric Field Instrument \citep{Wygant13}
and Electric and Magnetic Field Instrument Suite and Integrated Science \citep{Kletzing13} aboard the Van Allen Probe A on May 1, 2013. The waveforms are in the coordinate system related to the background (DC) magnetic field: $E_{\bot}$, $B_{\bot}$ are one of the electric and magnetic field components perpendicular to the background magnetic field, while $E_{\parallel}$, $B_{\parallel}$ are parallel to it. The middle panels present the waveform over 30 ms in the natural coordinate system shown in Figure \ref{fig2}. The right panel presents the spectrum of the electrostatic $E_z$ and electromagnetic $E_x$, $B_x$ fields for the selected 30 ms.\label{fig1}}
\end{figure*}

\begin{figure}
\includegraphics[width=11pc]{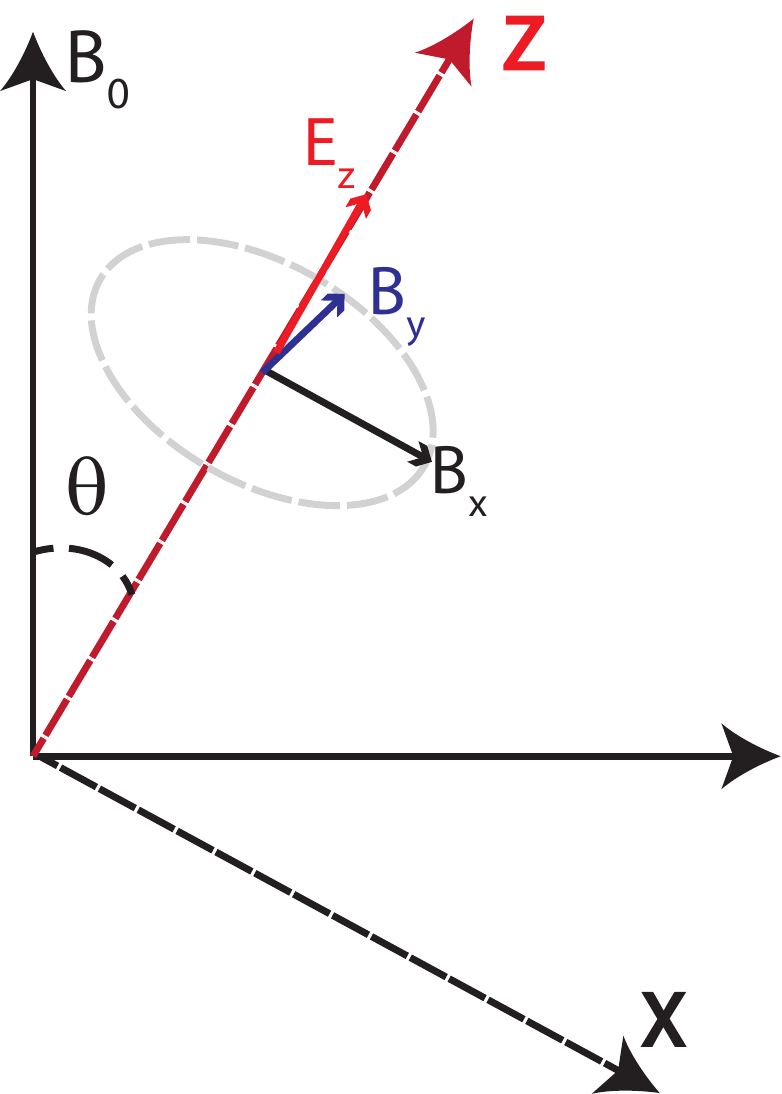}
\caption{The coordinate system with the $Z$ axis along the wave vector directed at wave normal angle $\theta$ with respect to the background magnetic field ${\bf B}_0$. The whistler wave field is decomposed into the electrostatic field $E_{z}$ and electromagnetic field in the $XY$ plane.\label{fig2}}
\end{figure}

Figure \ref{fig1} shows one second of the waveform of whistler waves continuously present for more than ten seconds around 11:27:25 UT. The left panels show the waveform in the coordinate system related to the background magnetic field. Surprisingly, the waveform of the parallel electric field is highly non-sinusoidal, in contrast to the other electric and magnetic field components. The middle panels present 30 ms of the waveform in the coordinate system with the $Z$ axis along the whistler wave vector that is the direction delivering minimum to the root mean square of ${\rm div}\;{\bf B}$ \citep{Sonnerup&Scheible98}. In this coordinate system the whistler wave field is decomposed into the electrostatic field $E_{z}$ and the electromagnetic field in the $XY$ plane (Fig. \ref{fig2}).

The selected 30 ms whistler wave packet propagates at the wave normal angle $\theta\sim 15^{\circ}$. The electromagnetic fields have quasi-sinusoidal waveforms and their spectra are peaked at $f\sim 450$ Hz that is about $0.2 f_c$. There is a good correlation between $E_y$ and $B_x$ in accordance with the Faraday's law, $E_y/B_x$ provides the phase velocity estimate of about $15000$ km/s that is in reasonable agreement with $c \left[f(f_c\cos\theta-f)/f_p^2\right]^{1/2}\sim 17500$ km/s from the dispersion relation in a cold plasma \citep{Helliwell65}. The electrostatic field $E_z$ has highly non-sinusoidal waveform with pronounced negative electric field spikes showing up in the wave spectrum as electric field power at harmonics of $\sim 450$ Hz. Van Allen Probe measurements show that the whistler waves are associated with the presence of the two-temperature electron population.

\begin{figure}
\includegraphics[width=16pc]{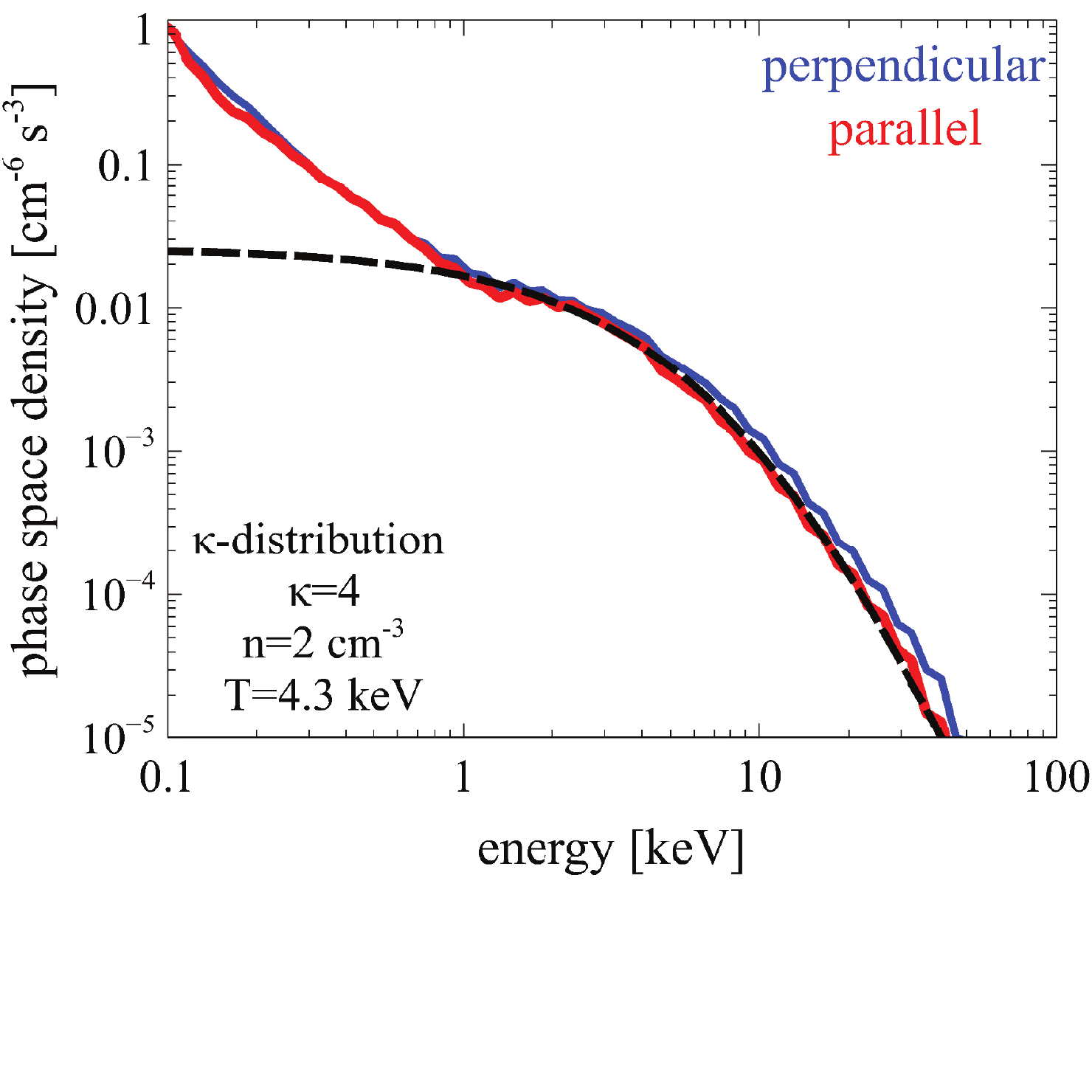}
\caption{The spectrum (phase space density) of electrons with pitch angles around 0$^{\circ}$ (streaming parallel to the background
magnetic field) and around 90$^{\circ}$ (streaming perpendicular to it) computed by converting the fluxes measured by the Helium Oxygen Proton Electron (HOPE) detector \citep{Funsten13} over 10 seconds around 11:27:25 UT. The high-energy part of the spectrum is fitted to $\mathcal{F}(\mathcal{E})=n\;\mathcal{C}_{\kappa}(m/2\pi \kappa \mathcal{E}_0)^{3/2}\left[1+\mathcal{E}/\kappa\mathcal{E}_0\right]^{-(\kappa+1)}$, where $\mathcal{C}_{\kappa}=\Gamma(\kappa+1)/\Gamma(\kappa-1/2)$, $n$ is the density and  $T=2\kappa\mathcal{E}_0/(2\kappa-3)$ is the temperature. The fitting parameters $\kappa$, $n$ and $T$ are presented in the panel. \label{fig3}}
\end{figure}

Figure \ref{fig3} presents cuts of the electron spectrum corresponding to fluxes of electrons with pitch angles around $0^{\circ}$ and $90^{\circ}$. The energy spectrum is anisotropic above a few keV as required to excite whistler waves via the cyclotron resonant instability \citep{Omura13:AGU}. The high-energy part of the electron spectrum is fitted to the $\kappa-$distribution resulting in density $\sim 2$ cm$^{-3}$ and temperature $\sim 4.3$ keV. The spectrum is not resolved below 100 eV because of contamination by photoelectrons. The total electron density of about 3.2 cm$^{-3}$ indicates the density of the low-energy electron population (below $\sim 1$ keV) of about 1.2 cm$^{-3}$. The temperature of the low-energy population is below a few hundred eV.

We address the whistler wave dynamics in the two-temperature electron plasma using the hydrodynamic and Maxwell equations. Because of the high whistler wave frequency ions can be considered as immobile charge neutralizing background \citep{Helliwell65}. In the coordinate system shown in Fig. \ref{fig2} the hydrodynamic equations for the two electron populations can be written as
\begin{eqnarray*}
\frac{d}{dt}\left[u_j-\frac{eA_{x}}{mc}\right]=-2\pi f_c\;v_j\cos\theta,\nonumber\\
\frac{d}{dt}\left[v_j-\frac{eA_{y}}{mc}\right]=2\pi f_c \left(u_j \cos\theta+w_j\sin\theta\right),
\end{eqnarray*}
\begin{eqnarray*}
\frac{dw_j}{dt}=\frac{e}{m}\frac{\partial \Phi}{\partial z}&-&\frac{1}{m\;n_j}\frac{\partial \left(T_j\;n_j\right)}{\partial z}-2\pi f_c\; v_j\sin\theta-\\&-&\frac{e}{mc}\left[u_j\frac{\partial A_x}{\partial z}+v_{j}\frac{\partial A_y}{\partial z}\right],
\end{eqnarray*}
\begin{eqnarray*}
\frac{dn_j}{dt}=-n_{j}\frac{\partial w_j}{\partial z},\;\;\;\frac{d}{dt}\equiv \frac{\partial}{\partial t}+w_{j}\frac{\partial}{\partial z},
\end{eqnarray*}
where $j=l,h$ corresponds to the low- and high-energy populations,  $n_j$ , $(u_j, v_j, w_j)$ and $T_j$ are electron densities, bulk velocities and temperatures, $-e$ and $m$ are electron charge and mass. The Maxwell equations for the electrostatic and vector potentials are ${\partial^2 \Phi}/{\partial z^2}=4\pi e\left(\sum_j n_j-n_0\right)$ and ${\partial^2{\bf A}}/{\partial z^2}=(4\pi e/c)\sum_j\left(n_ju_j\;\hat{x}+n_j v_j\;\hat{y}\right)$, where $n_0$ is the unperturbed electron density and the displacement current is neglected, because $\partial^2/\partial t^2\ll c^2\partial^2/\partial z^2$. We neglect the thermal spread of the low-energy population as not qualitatively affecting the whistler wave dynamics. The high-energy population is assumed to be isothermal $T_h={\rm const}$ as in the theory of ion-acoustic waves \citep{Sagdeev66}, because its thermal velocity is higher than the whistler wave phase velocity. By linearizing the equations we obtain the dispersion relations of linear waves propagating in the observed plasma.

\begin{figure}
\includegraphics[width=16pc]{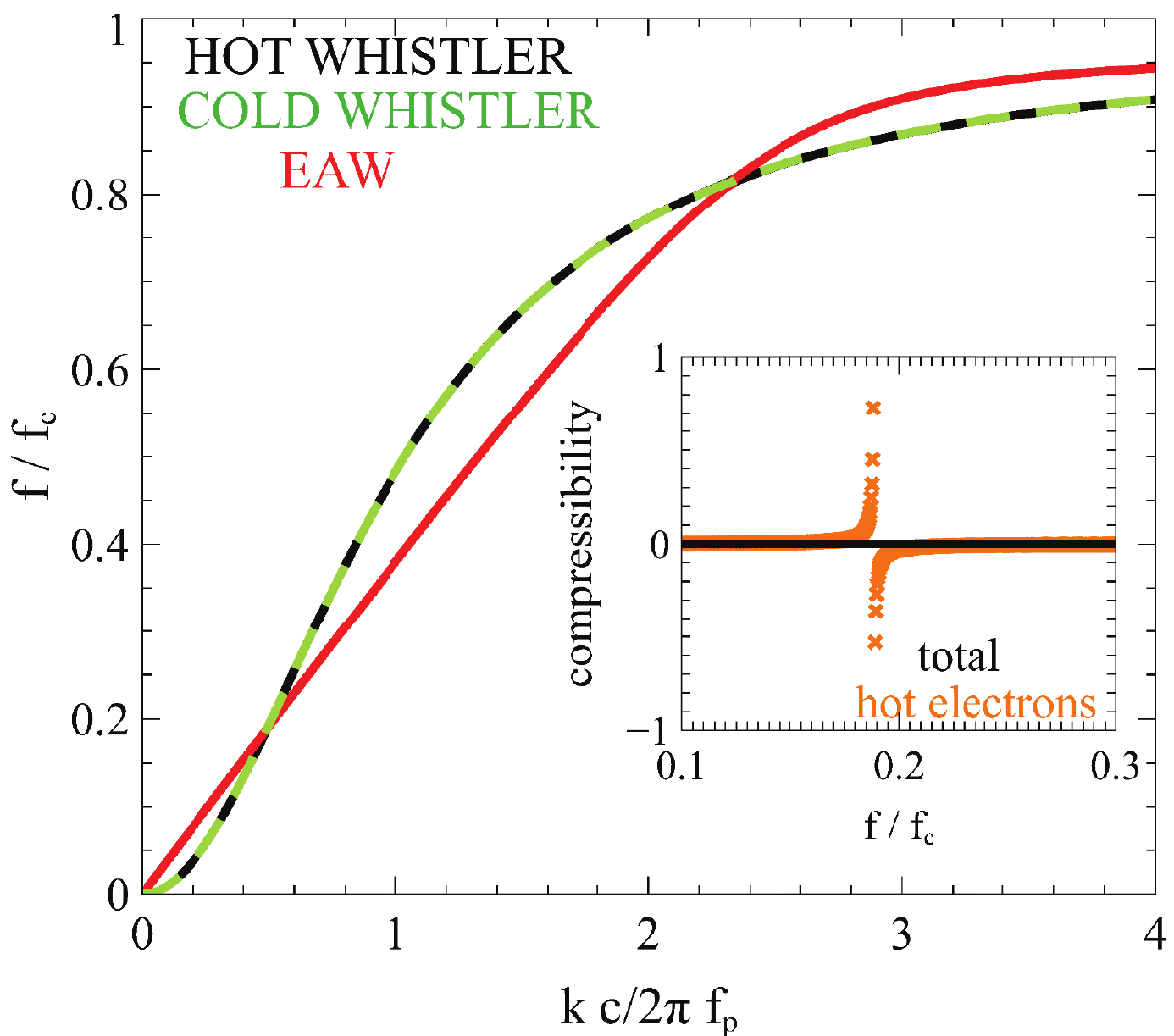}
\caption{The dispersion relation of linear waves propagating in the observed two-temperature electron plasma below the electron cyclotron frequency and above the low-hybrid frequency (ions are considered as immobile charge neutralizing background). The dispersion relations are computed for the wave normal angle $\theta=15^{\circ}$. The dashed green curve shows the dispersion relation of whistler waves in a cold plasma computed by setting zero thermal spread of the high-energy population. The small panel presents the compressibility of the high-energy population and the full electron compressibility defined as the ratio of amplitudes of the compressional velocities $w_h$ and $w_0$ and non-compressional bulk velocity $(u_0^2+v_0^2)^{1/2}$, where $(u_0,v_0,w_0)=\sum_j n_j (u_j,v_j,w_j)/n_0$ is the full electron bulk velocity. \label{fig4}}
\end{figure}

Figure \ref{fig4} shows that the whistler wave dispersion relation is not affected by the thermal spread and coincides with the cold dispersion relation, $f\approx f_c\cos\theta\;k^2c^2/(k^2c^2+4\pi^2f_p^2)$ \citep{Helliwell65}. The two-temperature electron plasma supports electrostatic electron-acoustic waves, which dispersion relation at long wavelengths is $2\pi f\approx k\;v_{EA}\cos\theta$, where $v_{EA}=(T_h/m)^{1/2}(n_c/n_0)^{1/2}$ is the electron-acoustic velocity \citep{Watanabe77,Gary&Tokar85}. Whistler and electron-acoustic waves turn out to be coupled in the crossover points at $f\sim 0.2 f_c$ and $\sim 0.8 f_c$. The most pronounced effect around the crossover frequencies is seen in the electron compressibility that is the ratio of amplitudes of compressional $w_j$ and non-compressional bulk velocities. The hot and cold electron populations become highly compressible around the crossover frequencies, although the full electron compressibility remains negligible.

We address the nonlinear evolution of whistler waves by solving the hydrodynamic and Maxwell equations using the energy conserving numerical scheme based on the Fast Fourier Transform previously used for analysis of steepening of electron-acoustic waves \citep{Dillard18}. The initial condition is a monochromatic whistler wave of a realistic amplitude. We have found that whistler waves with the frequency far from the crossover frequencies remain undisturbed in accordance with the previous simulations \citep{Yoon14}. On the contrary, whistler waves with the frequency around the crossover frequencies exhibit a fundamentally different behavior.

\begin{figure*}
\includegraphics[width=30pc]{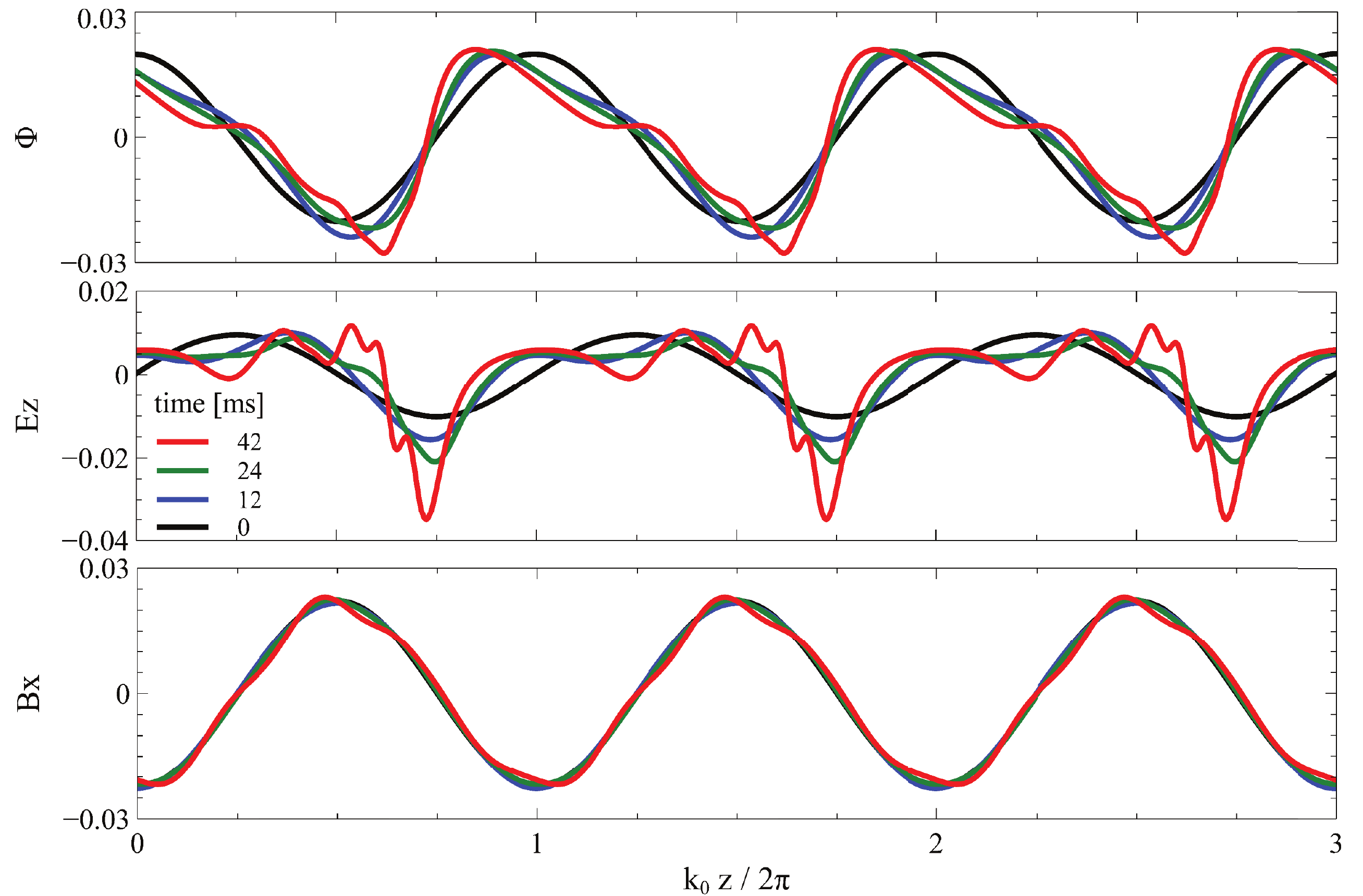}
\caption{The evolution of initially monochromatic whistler wave with a realistic amplitude and frequency $f_0$ around the first crossover frequency $\sim 0.2 f_c$. The simulation results are presented in the whistler wave reference frame. The electrostatic potential and electrostatic field are normalized to $m (cf_c\cos\theta/f_p)^2/e$ and $m (cf_c\cos\theta/f_p)^2/e d_e$, where $d_{e}=c/2\pi f_p$ is the electron inertial length. The magnetic field $B_x$ is normalized to the background magnetic field $B_0$. The electromagnetic field components $E_x, E_y$ and $B_y$ have profiles identical to $B_x$ up to a shift in phase. \label{fig5}}
\end{figure*}

Figure \ref{fig5} presents evolution of a whistler wave with the frequency $f_0$ around the first crossover frequency corresponding to the wave number $k_0\sim \pi f_p/c$. The evolution of the electrostatic potential exhibits signatures of the classical hydrodynamic steepening (overtaking) inherent to sound waves in fluids and plasmas \citep{Landau59,Sagdeev66}. The steepening produces the negative electric field spikes in the electrostatic component. The spikes become quite pronounced after about 40 ms and resemble those in observations. In accordance with observations, the magnetic field $B_{x}$ and the other electromagnetic components remain practically undisturbed.

The steepening of the whistler electrostatic field is identical to the steepening of electron-acoustic waves \citep{Dillard18}. In fact, the physics of the effect is equivalent. An initially monochromatic sufficiently long electron-acoustic wave $(f,k)$ produces harmonics $(\ell f,\ell k)$ that are in phase with the fundamental wave due to the linear dispersion relation at long wavelengths. Similarly, the whistler wave $(f_0,k_0)$ produces electron-acoustic waves $(\ell f_0, \ell k_0)$ that are in phase until $(\ell k_0)^{-1}$ becomes comparable to the dispersive scale of the electron-acoustic mode that is the Debye length $\lambda_D=(T_h/4\pi n_0 e^2)^{1/2}$. Therefore, the number of harmonics may not exceed $\ell\sim (k_0\lambda_D)^{-1}\sim c\;(T_h/m)^{-1/2}\sim 10$ that is consistent with the observed spectrum. Because the steepening of long electron-acoustic waves is known to be described by the Korteweg-de Vries equation \citep{Mace91}, we can estimate the steepening time of the whistler electrostatic field as $\tau_{s}\sim 2A(mT_h)^{1/2}\left(e E_0\right)^{-1}$, where $E_0$ is the initial amplitude of the whistler electrostatic field and $A^{-1}=(n_h/n_c)^{1/2}(3+n_c/n_h)$ \citep{Mace91,Dillard18}. Assuming the typical amplitude $E_0\sim 10$ mV/m  we find $\tau_{s}\sim 50$ ms that is consistent with the simulation results.

After about $60$ ms the energy leaks out of the steepening region due to the electron-acoustic wave dispersion converting the electric field spikes into oscillations that is inconsistent with observations. However, we have verified that inclusion of the collisional Burgers dissipation \citep{Landau59} does provide the persistency of the spikes. In collisionless plasma the dissipation is provided by the resonant wave-particle interaction. The wave-particle interaction effects were originally included into the hydrodynamic description of ion-acoustic waves \citep{Ott&Sudan69}. The recent simulations of electron-acoustic \citep{Dillard18} and Alfven \citep{Medvedev97} waves have confirmed that the wave-particle interaction does provide persistency of the electric field spikes and magnetic field pulses, respectively.

A whistler wave can efficiently exchange energy with electrons with the parallel velocity $v_{\parallel}$ satisfying the resonance condition, $\omega-k_{\parallel}v_{\parallel}=n \omega_c$, where $\omega=2\pi f$, $\omega_c=2\pi f_c$, $k_{\parallel}=k\cos\theta$ and $n=0,\pm 1,...$ \citep{Shklyar09}. The steepening of the whistler electrostatic field provides the wave energy cascade to higher frequencies and wave numbers opening the door for many more resonances: $\omega-k_{\parallel}v_{\parallel}=n\omega_{c}/\ell$. This facilitates the efficient energy mediation from the cyclotron resonant electrons, driving the whistler wave, to other electron populations, in particular, to lower energy electrons.

In summary, whistler waves propagating in the two-temperature electron plasma can exhibit the classical steepening producing pronounced spikes in the whistler electrostatic field. The steepening occurs due to whistler wave coupling to the electron-acoustic mode and becomes noticeable for whistler wave packets with non-negligible wave power at frequencies $f$ satisfying $4\pi n_c T_h/B_0^2\sim f/f_c\cos\theta-(f/f_c\cos\theta)^2$. The steepening explains surprising observations of whistler waves with significant electric field power at harmonics of the whistler wave fundamental frequency in the Van Allen radiation belts.

\begin{acknowledgments}
The work of I.V., O.A., F.M. and J.B. was performed under JHU/APL Contract No. 922613 (RBSP-EFW).
\end{acknowledgments}

%


\end{document}